\documentclass{elsart}

\usepackage{graphicx}
\usepackage{amssymb}

\begin{document}

\begin{frontmatter}



\title{Efficient atomic quantum memory for photonic qubits in cavity QED}


\author{Hiroyuki Yamada},
\author{Katsuji Yamamoto\corauthref{cor}}
\corauth[cor]{Corresponding author.}
\ead{yamamoto@nucleng.kyoto-u.ac.jp}

\address{Department of Nuclear Engineering,
Kyoto University, Kyoto 606-8501, Japan}

\begin{abstract}
We investigate a scheme of atomic quantum memory
to store photonic qubits of polarization in cavity QED.
It is observed that the quantum-state swapping
between a single-photon pulse and a $ \Lambda $-type atom
can be made via scattering in an optical cavity
[T.~W.~Chen, C.~K.~Law, P.~T.~Leung, Phys. Rev. A {\bf 69} (2004) 063810].
This swapping operates limitedly in the strong coupling regime
for $ \Lambda $-type atoms with equal dipole couplings.
We extend this scheme in cavity QED
to present a more feasible and efficient method
for quantum memory combined with projective measurement.
This method works without requiring such a condition
on the dipole couplings.
The fidelity is significantly higher than that of the swapping,
and even in the moderate coupling regime
it reaches almost unity by narrowing sufficiently the photon-pulse spectrum.
This high performance is rather unaffected
by the atomic loss, cavity leakage or detunings,
while a trade-off is paid in the success probability
for projective measurement.
\end{abstract}

\end{frontmatter}


\section{Introduction}
\label{sec:introduction}

Combined systems of atoms and photons have been studied extensively
to construct promising and efficient quantum networks
for information processing and communication
\cite{qnet}.
In these quantum networks quantum-state transfer
between photons and atoms (matter),
and storage of quantum data are particularly important.
Then, numerous methods to implement the quantum-state transfer
and quantum memory have been proposed and investigated
in various manners
\cite{qtrqm-1,qtrqm-2,qtrqm-3,qtrqm-4,qtrqm-5,qtrqm-6,qtrqm-7,qtrqm-8,qtrqm-9,qtrqm-10,qtrqm-11,qtrqm-12,qtrqm-13,qtrqm-14,qtrqm-15,qtrqm-16,qtrqm-17,qtrqm-18,qtrqm-19}.
The cavity QED is among the promising schemes
to realize such quantum-state operations,
which utilizes strong interactions between single atoms and photons
inside cavities
\cite{CQED,atom-CQED}.
Specifically, quantum-state transfer and manipulation
are made between a single atom and a single-photon pulse
through a scattering in an optical cavity.

In this paper we investigate a scheme of atomic quantum memory
to store photonic qubits of polarization in cavity QED.
It is observed that the quantum-state swapping
between a single-photon pulse and a $ \Lambda $-type atom
can be made via scattering in an optical cavity
\cite{CLL-2004}.
This swapping operates limitedly in the strong coupling regime
for $ \Lambda $-type atoms with equal dipole couplings.
We extend this scheme in cavity QED
to present a more feasible and efficient method
for quantum memory operation,
storage and retrieval of photonic qubits,
combined with projective measurement.

The present method for quantum memory
has some characteristic properties and advantages as follows.
First, quantum information is encoded as qubits
in polarization states of single-photon pulses.
Such photonic qubits are then stored
in the two degenerate stable ground states
(e.g., Zeeman sublevels of a hyperfine ground state)
of single atoms trapped in optical cavities.
Hence, this atomic quantum memory is expected
to be robust against decoherence.
The memory operation is performed manifestly
for atomic and photonic qubits via scatterings,
which is suitable for the discrete-variable quantum information scheme.

Since the quantum-state transformation for memory operation
is made on the asymptotic states after scattering,
no precise timing of the interaction period is required
\cite{CLL-2004}.
Furthermore, no control light is used
for the quantum-state transfer
in contrast with many other proposals for quantum memory.
The present method hence operates in a passive way
except for the projective measurement.
This would be favorable for scalability.

The naive quantum-state swapping via atom-photon scattering
operates in rather limited situations;
$ \Lambda $-type atoms with equal dipole couplings,
and the strong coupling regime with negligible detunings
\cite{CLL-2004}.
In contrast, the present quantum memory operation
combined with projective measurement
can be implemented in more practical situations.
It works without requiring such a condition
on the dipole couplings of $ \Lambda $-type atoms.
The fidelity is significantly higher than that of the swapping,
and even in the moderate coupling regime
it reaches almost unity by narrowing sufficiently the photon-pulse spectrum.
This high performance is rather unaffected
by the atomic loss, cavity leakage or detunings,
while a trade-off is paid in the success probability
for projective measurement.

The rest of the paper is organized as follows.
In Sec. \ref{sec:scattering}, the basic description
of the atom-photon scattering in cavity QED is reviewed,
and the swapping between atomic and photonic qubits via scattering
is discussed.
In Sec. \ref{sec:memory-operation},
by extending this scheme of quantum-state transfer in cavity QED,
we present a more feasible and efficient method
for quantum memory operation combined with projective measurement.
A numerical analysis is presented in Sec. \ref{sec:efficiency}
to exhibit the high performance of the present atomic quantum memory.
In Sec. \ref{sec:2-qubit}, storage of 2-qubit photonic entanglement
is also considered as an application.
Section \ref{sec:summary} is devoted to summary.

\section{Atom-photon scattering
and quantum-state transformation in cavity QED}
\label{sec:scattering}

We first review the basic description
of the atom-photon scattering and quantum-state transformation
in cavity QED \cite{CLL-2004}.
A one-dimensional cavity bounded by two mirrors is used,
one of which is perfectly reflecting
while the other is partially transparent.
The electromagnetic field is expanded in terms of the continuous modes
with the wave number $ k $, which range over the inside of cavity
through the outside free space
\cite{cpm}.
A photonic qubit is encoded in the polarization states
$ | k_L \rangle $ and $ | k_R \rangle $ of single-photon pulse as
\begin{eqnarray}
| \phi_{\rm p} \rangle
&=& c_L | {\bar k}_L \rangle + c_R | {\bar k}_R \rangle
= \int\nolimits_{- \infty}^{\infty} dk f(k) e^{-ikt}
| \phi_{\rm p} (k) \rangle ,
\label{eqn:phi-p}
\\
| {\bar k}_{L,R} \rangle
&=& \int\nolimits_{- \infty}^{\infty} dk f(k) e^{-ikt} | k_{L,R} \rangle ,
\end{eqnarray}
where $ f(k) $ is the normalized spectral amplitude,
and $ e^{-ikt} $ represents the asymptotic temporal evolution
($ c = \hbar = 1 $ unit).
On the other hand, a $ \Lambda $-type atom is trapped
inside the cavity, which has two degenerate ground states
$ | L \rangle $ and $ | R \rangle $ and an excited state $ | e \rangle $.
Then, an atomic qubit is encoded in the degenerate ground states as
\begin{eqnarray}
| \psi_{\rm a} \rangle = a_L | L \rangle + a_R | R \rangle .
\label{eqn:psi-a}
\end{eqnarray}

The polarization states $ | k_L \rangle $ and $ | k_R \rangle $
are coupled, respectively, to the transitions
$ | L \rangle - | e \rangle $ and $ | R \rangle - | e \rangle $
of a frequency $ \omega_e $ in the cavity with dipole couplings
\begin{eqnarray}
g_{L,R} (k) = \frac{\lambda_{L,R} {\sqrt{\kappa / \pi}}e^{i \theta_{L,R}}}
{k - k_c + i \kappa } ,
\label{eqn:gLR}
\end{eqnarray}
where $ k_c $ is the resonant frequency of the cavity,
$ \kappa $ is the leakage rate of the cavity,
$ \lambda_L $ and $ \lambda_R $ represent the normalized coupling strengths
(single-photon Rabi frequencies),
and $ \theta_L $ and $ \theta_R $ are the phase angles
from the dipole transition matrix elements.
(It will be seen later that the phase angles
$ \theta_L $ and $ \theta_R $ are irrelevant
for the transfer between atomic and photonic qubits via scattering.
Actually, they may be removed at the beginning
by phase transformations of $ | L \rangle $ and $ | R \rangle $.)
The atom-photon scattering takes place through these couplings,
and the transformation of the atom and photon states is made
asymptotically as
\begin{eqnarray}
{\mathcal T} | L k_L \rangle
&=& T_{LL} (k) | L k_L \rangle + T_{RL} (k) | R k_R \rangle ,
\nonumber \\
{\mathcal T} | R k_R \rangle
&=& T_{LR} (k) | L k_L \rangle + T_{RR} (k) | R k_R \rangle ,
\nonumber \\
{\mathcal T} | L k_R \rangle
&=& | L k_R \rangle ,
\nonumber \\
{\mathcal T} | R k_L \rangle
&=& | R k_L \rangle ,
\label{eqn:transform}
\end{eqnarray}
where the basis states are taken as
$ | L k_L \rangle \equiv | L \rangle | k_L \rangle $ and so on.
The scattering matrix elements are calculated \cite{CLL-2004} as
\begin{eqnarray}
&& T_{LL} (k) = e^{i \phi_s (k)} \sin^2 \xi^2 + \cos^2 \xi ,
\nonumber \\
&& T_{RR} (k) = \sin^2 \xi + e^{i \phi_s (k)} \cos^2 \xi ,
\nonumber \\
&& T_{LR} (k) = e^{- i ( \theta_L - \theta_R )} \sin \xi \cos \xi
( e^{i \phi_s (k)} - 1 ) ,
\nonumber \\
&& T_{RL} (k) = e^{i ( \theta_L - \theta_R )} \sin \xi \cos \xi
( e^{i \phi_s (k)} - 1 ) ,
\label{eqn:T-matrix}
\end{eqnarray}
where
\begin{eqnarray}
\sin \xi &=& \lambda_L / \lambda , \
\cos \xi = \lambda_R / \lambda ,
\\
\lambda &=& \sqrt{\lambda_L^2 + \lambda_R^2} .
\end{eqnarray}
The phase shift $ \phi_s(k) $ acquired by the bright state
is determined independently of the photon-pulse shape $ f(k) $.
It is explicitly calculated as
\begin{eqnarray}
e^{i \phi_s (k)}
= \frac{( k - k_c + i \kappa ) w_+ ( k - k_c )}
{( k - k_c - i \kappa ) w_- ( k - k_c )} ,
\label{eqn:phi-s}
\end{eqnarray}
where
\begin{eqnarray}
w_\pm (s)
= s^2 - ( \delta_e - i \gamma \pm i \kappa ) s
- \lambda^2 \pm i \kappa ( \delta_e - i \gamma )
\end{eqnarray}
with the atomic detuning $ \delta_e = \omega_e - k_c $.
The linear transformation $ {\mathcal T} $ via scattering
with the complex $ \phi_s(k) $ is generally non-unitary
due to the atomic loss with a rate $ \gamma $
induced by the spontaneous emission into the environment.

It is observed in Ref. \cite{CLL-2004}
that the quantum-state swapping between the atom and photon
can be made via scattering in the strong coupling regime
for a $ \Lambda $-type atom having equal dipole couplings,
\begin{eqnarray}
\lambda_L = \lambda_R = \lambda /{\sqrt 2} .
\label{eqn:lmL-lmR}
\end{eqnarray}
(This codition is met, for example, by the D1 line of sodium with
$ | L \rangle = | F = 1 , m_F = -1 \rangle $,
$ | R \rangle = | F = 1 , m_F = 1 \rangle $,
$ | e \rangle = | F^\prime = 1 , m_{F^\prime} = 0 \rangle $.)
In fact, with the maximal phase shift
$ e^{i \phi_s ( k_c )} = - 1 $ at the resonance
($ k = k_c = \omega_e $) in the strong coupling limit
$ \lambda^2 / \kappa \gamma \rightarrow \infty $,
we have the scattering matrix elements as
$ e^{i ( \theta_L - \theta_R )} T_{LR}( k_c )
= e^{- i ( \theta_L - \theta_R )} T_{RL} ( k_c ) = - 1 $,
$ T_{LL}( k_c ) = T_{RR}( k_c ) = 0 $ in Eq. (\ref{eqn:T-matrix})
under the condition $ \lambda_L = \lambda_R $.
Then, the swapping between the atomic and photonic qubits is made as
\begin{eqnarray}
\begin{array}{c}
| \psi_{\rm a} \rangle | \phi_{\rm p} ( k_c ) \rangle
\stackrel{\mathcal T}{\Rightarrow}
e^{- i ( \theta_L + \theta_R )}
| \psi_{\rm swap} \rangle | \phi_{\rm swap} ( k_c ) \rangle ,
\end{array}
\end{eqnarray}
where
\begin{eqnarray}
| \psi_{\rm swap} \rangle
&=& c_R ( e^{i \theta_R} | L \rangle )
+ c_L ( - e^{i \theta_L} | R \rangle ) ,
\label{eqn:psi-swap}
\\
| \phi_{\rm swap} (k) \rangle
&=& a_R ( - e^{i \theta_R} | k_L \rangle )
+ a_L ( e^{i \theta_L} | k_R \rangle ) .
\label{eqn:phi-swap}
\end{eqnarray}

It may be expected naively that this swapping is applicable
for storing photonic qubits in atomic qubits.
In order to examine the feasibility of atom-photon swapping
for quantum memory we evaluate the fidelity as follows.
Arbitrary atomic and photonic qubits
in Eqs. (\ref{eqn:phi-p}) and (\ref{eqn:psi-a})
are taken as the initial state
\begin{eqnarray}
| \Phi_{\rm in} \rangle = | \psi_{\rm a} \rangle | \phi_{\rm p} \rangle .
\end{eqnarray}
Then, the density operator of the output state via scattering
is given by
\begin{eqnarray}
\rho_{\rm out} = | \Phi_{\rm out} \rangle \langle \Phi_{\rm out} |
+ ( 1 - \langle \Phi_{\rm out} | \Phi_{\rm out} \rangle )
| 0 \rangle \langle 0 |
\label{eqn:rho-out}
\end{eqnarray}
with
\begin{eqnarray}
| \Phi_{\rm out} \rangle
&=& {\mathcal T} | \Phi_{\rm in} \rangle
= \int\nolimits_{- \infty}^{\infty} dk f(k) e^{-ikt}
| \Phi_{\rm out} (k) \rangle ,
\\
| \Phi_{\rm out} (k) \rangle
&=& - e^{- 2i \theta_R}
T_{LR}(k) | \psi_{\rm swap} \rangle | \phi_{\rm swap} (k) \rangle
\nonumber \\
&{}& + T_{LL}(k) | \psi_{\rm a} \rangle | \phi_{\rm p} (k) \rangle .
\end{eqnarray}
Here, the relations
$ T_{LL}(k) = T_{RR}(k) $,
$ e^{i ( \theta_L - \theta_R )} T_{LR}(k)
= e^{- i ( \theta_L - \theta_R )} T_{RL}(k) $
and $ T_{LL}(k) - e^{i ( \theta_L - \theta_R )} T_{LR}(k) = 1 $
under the condition $ \lambda_L = \lambda_R $ are considered.
The ``vacuum" term of $ | 0 \rangle \langle 0 | $
represents the loss due to the spontaneous emission,
providing $ {\rm Tr} \rho_{\rm out} = 1 $.
The output photon will eventually be absorbed by matter.
Then, by taking the trace over the photon states
the fidelity to obtain the desired atomic state
$ | \psi_{\rm swap} \rangle $ is given by
\begin{eqnarray}
F ( | \phi_{\rm p} \rangle )
= \left[ \langle \psi_{\rm swap} | {\rm Tr}_{\rm p}
\left[ | \Phi_{\rm out} (k) \rangle \langle \Phi_{\rm out} (k) | \right]
| \psi_{\rm swap} \rangle \right]_f ,
\label{eqn:F-pa}
\end{eqnarray}
where $ {\rm Tr}_{\rm p} [ \rho ] \equiv \langle k_L | \rho | k_L \rangle
+ \langle k_R | \rho | k_R \rangle $,
and the average of any function $ G(k) $ of $ k $
with the weight $ | f(k) |^2 $ is denoted as
\begin{eqnarray}
[ G(k) ]_f \equiv \int\nolimits_{- \infty}^{\infty} dk | f(k) |^2 G(k) .
\end{eqnarray}
This fidelity is calculated by considering the relation
$ T_{LL}(k) - e^{i ( \theta_L - \theta_R )} T_{LR}(k) = 1 $
under the condition $ \lambda_L = \lambda_R $ as
\begin{eqnarray}
F ( | \phi_{\rm p} \rangle )
= F_{\rm swap} + ( 1 -  F_{\rm swap} )
| \langle \psi_{\rm swap} | \psi_{\rm a} \rangle |^2
\geq F_{\rm swap} .
\label{eqn:F-pa-F-swap}
\end{eqnarray}
The fidelity of swapping is then given
irrespective of the choice of initial state as
\begin{eqnarray}
F_{\rm swap} = \left[ | \sin ( \phi_s (k) / 2 ) |^2 \right]_f
( \lambda_L = \lambda_R ) .
\label{eqn:F-swap}
\end{eqnarray}
Here, it should be noted that this fidelity of swapping $ F_{\rm swap} $
is meaningful only for the case $ \lambda_L = \lambda_R $
in calculating $ F ( | \phi_{\rm p} \rangle ) $
as Eq. (\ref{eqn:F-pa-F-swap})
with $ | T_{LR}(k) | = | \sin ( \phi_s (k) / 2 ) | $.

The atom-photon swapping may be optimized
by satisfying the conditions
$ k_p = k_c = \omega_e $, $ \kappa_p \ll \kappa $
and $ \lambda^2 / \kappa \gamma \gg 1 $,
as discussed in Ref. \cite{CLL-2004},
where $ k_p $ and $ \kappa_p $ represent
the peak position and width of the photon-pulse spectrum $ f(k) $,
respectively.
Specifically, by including the effects of detunings
$ \delta_e = \omega_e - k_c $ and $ \delta_p = k_p - k_c $
the fidelity of swapping $ F_{\rm swap} $ in Eq. (\ref{eqn:F-swap})
is evaluated in the leading terms for $ \kappa_p \rightarrow 0 $ as
\begin{eqnarray}
F_{\rm swap} & \approx & 1 - 2 ( \kappa \gamma / \lambda^2 )
- [ ( \kappa \delta_e / \lambda^2 ) + ( \delta_p / \kappa ) ]^2 .
\label{eqn:F-swap-leading}
\end{eqnarray}
In order to obtain $ F_{\rm swap} \geq 0.99 $, for instance,
we need to arrange roughly
$ \kappa \gamma / \lambda^2 \lesssim 1/200 $,
$ \kappa | \delta_e | / \lambda^2 \lesssim 1/10 $
and $ | \delta_p | / \kappa \lesssim 1/10 $
unless a tuning is made as
$ \delta_p \approx - ( \kappa / \lambda )^2 \delta_e $.
Hence, a rather strong atom-photon coupling is required
to achieve a high fidelity.
It should be mentioned further
that this atom-photon swapping operates
under the condition $ \lambda_L = \lambda_R $
on the dipole couplings.
In general, for $ \lambda_L \not= \lambda_R $
the transfer between atomic and photonic qubits
is not implemented sufficiently even in the strong coupling limit
$ \lambda^2 / \kappa \gamma \rightarrow \infty $.
We can check specifically the relation
\begin{eqnarray}
F ( | {\bar k}_L \rangle ) + F ( | {\bar k}_R \rangle )
= 1 + \left[ | T_{LR}(k) |^2 \right]_f \leq 1 + \sin^2 2 \xi
\end{eqnarray}
by calculating the fidelity in Eq. (\ref{eqn:F-pa})
for $ | {\bar k}_L \rangle $ and $ | {\bar k}_R \rangle $
as the initial photonic qubit $ | \phi_{\rm p} \rangle $.
This indicates that if $ \lambda_L \not= \lambda_R $
($ \sin 2 \xi < 1 $) the fidelity of qubit transfer
$ F ( | \phi_{\rm p} \rangle ) $ remains rather below unity
for some set of qubits,
which is insufficient for quantum memory to store unknown qubits.

\section{Memory operation: storage and retrieval of photonic qubits
with projective measurements}
\label{sec:memory-operation}

We here present an extended scheme for quantum memory
via scattering in cavity QED, which works more efficiently
with high fidelity in practical situations.
We consider the sequence of storage and retrieval of photonic qubits,
which may appear somewhat different from simply repeating
twice the atom-photon swapping.
To be general, we do not assume the condition (\ref{eqn:lmL-lmR})
for $ \Lambda $-type atoms, allowing different dipole couplings.
Then, quantum memory operation can be implemented
conditionally by making some projective measurements,
as described in detail below.
It will be seen that the fidelity of almost unity is achieved
by narrowing sufficiently the spectral width of the photon pulse
($ \kappa_p \ll \kappa $).

\begin{figure}[t]
\begin{center}
\scalebox{1.25}{\includegraphics*[1.5cm,1.5cm][8.5cm,6.5cm]{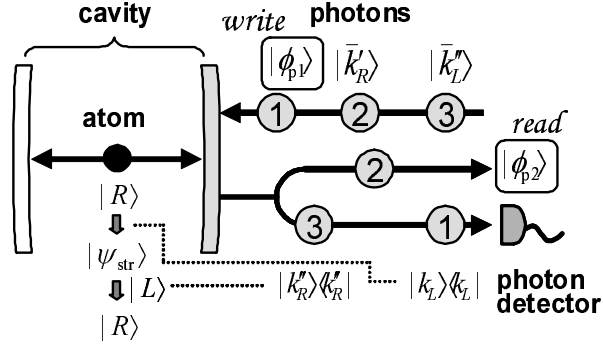}}
\caption{The schematic diagram
for the sequence of storage and retrieval of a photonic qubit
via scattering combined with photon-polarization measurements.
The qubit encoded in the first photon pulse is stored in the atom
by detecting the output photon.
Then, the second photon pulse is injected to retrieve the stored qubit.
The third photon pulse is used in the retrieval process
to project out the $ | L \rangle $ component of the atomic state
as recovering the initial $ | R \rangle $ state.
}
\label{scheme}
\end{center}
\end{figure}

The schematic diagram for the sequence of storage and retrieval
of a photonic qubit is depicted in Fig. \ref{scheme}.
The initial state is taken specifically as
\begin{eqnarray}
| \Phi_{\rm in} \rangle
= | R \rangle | \phi_{\rm p1} \rangle | {\bar k}_R^\prime \rangle
\label{eqn:Phi-in}
\end{eqnarray}
with
\begin{eqnarray}
| \phi_{\rm p1} \rangle
= c_L | {\bar k}_L \rangle + c_R | {\bar k}_R \rangle .
\end{eqnarray}
(The third photon of $ | {\bar k}_L^{\prime \prime} \rangle $
is omitted for simplicity.)
Here, $ | \phi_{\rm p1} \rangle $ is an unknown photonic qubit
to be stored and then retrieved.
The atomic state is initially prepared to be $ | R \rangle $,
and the second photon pulse of $ | {\bar k}_R^\prime \rangle $
is injected after a time delay $ \tau $
($ \gg \kappa_p^{-1} \gg \kappa^{-1} $) to retrieve the stored qubit.
We take for definiteness the same profile $ f(k) $
for the photon pulses (though this need not be required precisely).

{\mbox{}}

\begin{flushleft}
{\it Storage:}
\end{flushleft}

In the storage process, after the first photon pulse is scattered
with the atom, the polarization ``$ L $" is detected on the output photon.
This polarization detection is represented
by a positive operator-valued measure
\begin{eqnarray}
\Pi ( k_L ) = \int\nolimits_{- \infty}^{\infty} dk \eta (k)
| k_L \rangle \langle k_L |
\end{eqnarray}
with the quantum efficiency $ 0 < \eta (k) \leq 1 $.
[The dark count is neglected here
since it can actually be made rather small.
The terms of more than one photon states
may also be discarded effectively in $ \Pi ( k_L ) $
for the present process involving a single atom and a single photon.]
Then, the resultant state is given by
\begin{eqnarray}
\rho_1 &=& \frac{1}{P( k_L )} {\rm Tr}_{\rm p1} [ \Pi ( k_L )
{\mathcal T}_1 | \Phi_{\rm in} \rangle
\langle \Phi_{\rm in} | {\mathcal T}_1^\dagger ]
\nonumber \\
&=& \int\nolimits_{- \infty}^{\infty} dk
\frac{\eta (k) | f(k) |^2}{P( k_L )}
| \psi_{\rm str} (k) \rangle \langle \psi_{\rm str} (k) |
\otimes | {\bar k}_R^\prime \rangle \langle {\bar k}_R^\prime | ,
\end{eqnarray}
where by applying Eq. (\ref{eqn:transform}) for $ {\mathcal T}_1 $,
\begin{eqnarray}
| \psi_{\rm str} (k) \rangle
= T_{LR}(k) c_R | L \rangle + c_L | R \rangle .
\label{eqn:psi-a-str}
\end{eqnarray}
The success probability of this operation is also calculated
with $ {\rm Tr}_{\rm ap2} \rho_1 = 1 $ as
\begin{eqnarray}
P( k_L ) = \left[ \eta (k)
\langle \psi_{\rm str} (k) | \psi_{\rm str} (k) \rangle
\right]_f .
\label{eqn:P-kL}
\end{eqnarray}

It is noticed in Eq. (\ref{eqn:psi-a-str}) that the initial photonic qubit
is transferred to the atomic qubit with slight modification
by the factor $ T_{LR}(k) $.
This quantum-state transfer between the atom and photon
combined with projective measurement may be viewed
as a sort of one-bit teleportation \cite{OBT},
where the scattering acts as a 2-qubit gate
to create the entanglement of atom and photon.
The loss term of $ | 0 \rangle \langle 0 | $,
as seen in Eq. (\ref{eqn:rho-out}),
is removed by the photon detection even with $ \eta (k) < 1 $
(and the negligible dark count).
In the ideal case of strong coupling limit with $ \lambda_L = \lambda_R $
the initial atomic state $ | R \rangle $
in Eq. (\ref{eqn:Phi-in}) is swapped entirely
to the left-polarized photon $ | k_L \rangle $
in the output state via scattering, as discussed in the preceding section.
[See Eqs. (\ref{eqn:psi-a}) and (\ref{eqn:phi-swap})
with $ a_L = 0 $ and $ a_R = 1 $.]
Then, the probability for the detection of left-polarized photon
$ P( k_L ) $ in Eq. (\ref{eqn:P-kL}) becomes unity
with the full quantum efficiency.

{\mbox{}}

\begin{flushleft}
{\it Retrieval:}
\end{flushleft}

The retrieval of the photonic qubit is implemented
by the scattering of the second photon pulse
followed by the detection of the atomic state $ | L \rangle $.
The resultant output state is given by
\begin{eqnarray}
\rho_2
&=& \frac{1}{P(L)}
\langle L | {\mathcal T}_2 \rho_1 {\mathcal T}_2^\dagger | L \rangle
\nonumber \\
&=&
\frac{| T_{LR}( k_p ) |^2}{P( k_L ) P(L)}
\int\nolimits_{- \infty}^{\infty} dk
\eta (k) | f(k) |^2 
| \phi_{\rm rtr} (k) \rangle \langle \phi_{\rm rtr} (k) |
\label{eqn:rho-2}
\end{eqnarray}
with
\begin{eqnarray}
| \phi_{\rm rtr} (k) \rangle
&=& \int\nolimits_{- \infty}^{\infty} dk^\prime
f( k^\prime ) e^{-ik^\prime ( t - \tau )}
\nonumber \\
&{}& \times \frac{[ T_{LR}(k) c_R | k^\prime_R \rangle
+ T_{LR}( k^\prime ) c_L | k^\prime_L \rangle ]}
{T_{LR}( k_p )} ,
\label{eqn:phi-rtr}
\end{eqnarray}
where $ {\rm Tr}_{\rm p2} \rho_2 = 1 $ to determine
the probability $ P(L) $ for the projective measurement of $ | L \rangle $.
It is found here that for $ T_{LR}(k) \approx T_{LR}( k_p ) $
in the vicinity of narrow pulse peak
$ | k - k_p | \lesssim \kappa_p \ll \kappa $,
the output state becomes very closed to the desired photon state
as retrieval:
\begin{eqnarray}
| \phi_{\rm rtr} (k) \rangle \approx | \phi_{\rm p2} \rangle
= c_L | {\bar k}_L^\prime \rangle + c_R | {\bar k}_R^\prime \rangle .
\label{eqn:phi-p2}
\end{eqnarray}
Although some modification is made on the stored atomic qubit,
as seen in Eq. (\ref{eqn:psi-a-str}),
it is nearly compensated by the retrieval process,
as seen in Eq. (\ref{eqn:phi-p2}),
hence realizing the almost faithful retrieval of the initial photonic qubit.
A trade-off for this high fidelity should be paid
rather in the success probability,
as will be seen in Eq. (\ref{eqn:P-qm}).

{\mbox{}}

\begin{flushleft}
{\it Fidelity and success probability:}
\end{flushleft}

The quantum memory operation described so far is summarized
(see Fig. \ref{scheme}) as
\begin{eqnarray}
\begin{array}{c}
| \phi_{\rm p1} \rangle
= c_L | {\bar k}_L \rangle + c_R | {\bar k}_R \rangle
\\
{\mathcal T}_1 \Downarrow \Pi ( k_L )
\\
| \psi_{\rm str} (k) \rangle
= T_{LR}(k) c_R | L \rangle + c_L | R \rangle
\\
{\mathcal T}_2 \Downarrow | L \rangle \langle L |
\\
| \phi_{\rm rtr} (k) \rangle \approx | \phi_{\rm p2} \rangle
= c_L | {\bar k}_L^\prime \rangle + c_R | {\bar k}_R^\prime \rangle .
\end{array}
\end{eqnarray}
The fidelity for this sequence of storage and retrieval
is evaluated from Eqs. (\ref{eqn:rho-2}) and (\ref{eqn:phi-rtr}) as
\begin{eqnarray}
F ( {\rm p1} \rightarrow {\rm a} \rightarrow {\rm p2} )
&=& \langle \phi_{\rm p2} | \rho_2 | \phi_{\rm p2} \rangle
\nonumber \\
&=& \frac{\left[ \eta (k)
| \langle \phi_{\rm p2} | \phi_{\rm rtr} (k) \rangle |^2
\right]_f}
{\left[ \eta (k)
\langle \phi_{\rm rtr} (k) | \phi_{\rm rtr} (k) \rangle
\right]_f} .
\label{eqn:F-p1ap2}
\end{eqnarray}
Since the quantum efficiency may be taken as constant,
$ \eta (k) = \eta $, around the pulse peak in a good approximation,
it is actually calculated in Eq. (\ref{eqn:F-p1ap2}) as
\begin{eqnarray}
F ( {\rm p1} \rightarrow {\rm a} \rightarrow {\rm p2} )
= F_{\rm qm} + ( 1 - F_{\rm qm} ) | c_L |^4 \geq F_{\rm qm}
\end{eqnarray}
depending on $ | \phi_{\rm p1} \rangle $.
The fidelity of quantum memory is then given by
\begin{eqnarray}
F_{\rm qm}
&=& | \left[ T_{LR} (k) \right]_f |^2 / \left[ | T_{LR} (k) |^2 \right]_f
\nonumber \\
&=& \frac{| \left[ \sin ( \phi_s (k) / 2 ) \right]_f |^2}
{\left[ | \sin ( \phi_s (k) / 2 ) |^2 \right]_f}
\label{eqn:F-qm}
\end{eqnarray}
irrespective of the choice of unknown initial state.
This fidelity $ F_{\rm qm} $ is independent
of the ratio $ \lambda_L / \lambda_R $.
Hence, the present quantum memory works
without relying on the specific relation of the dipole couplings.
The net success probability of quantum memory is also calculated
from Eq. (\ref{eqn:rho-2}) with $ {\rm Tr}_{\rm p2} \rho_2 = 1 $ as
\begin{eqnarray}
P_{\rm qm} &=& P( k_L ) P(L)
= \eta \left[ | T_{LR} (k) |^2 \right]_f
= \eta \sin^2 2 \xi F_{\rm swap} ,
\label{eqn:P-qm}
\end{eqnarray}
where $ \sin 2 \xi = 2 ( \lambda_L / \lambda ) ( \lambda_R / \lambda ) $
depending on $ \lambda_L / \lambda_R $.
It is noticed here that $ P_{\rm qm} $ is intimately related to
$ F_{\rm swap}
= \left[ | \sin ( \phi_s (k) / 2 ) |^2 \right]_f $
given in Eq. (\ref{eqn:F-swap}).

As explained for $ P( k_L ) $ in the storage process,
the second photon pulse $ | {\bar k}_R^\prime \rangle $
in Eq. (\ref{eqn:Phi-in}) is swapped ideally
to the atomic state $ | L \rangle $ with $ P(L) = 1 $
in the strong coupling limit.
Then, the success probability of quantum memory $ P_{\rm qm} $
becomes unity together with the fidelity of swapping $ F_{\rm swap} $
in Eq. (\ref{eqn:P-qm})
with $ \eta = 1 $ and $ \sin 2 \xi = 1 $ ($ \lambda_L = \lambda_R $).
The atomic detection of $ | L \rangle $ in Eq. (\ref{eqn:rho-2})
for the retrieval process may be implemented
by using the third photon pulse.
Specifically, by injecting the photon pulse
of $ | {\bar k_L}^{\prime \prime} \rangle $ into the cavity,
the $ | L \rangle | k_L^{\prime \prime} \rangle $ component
is transformed to
$ T_{RL} ( k^{\prime \prime} )
| R \rangle | k_R^{\prime \prime} \rangle
+ T_{LL} ( k^{\prime \prime} )
| L \rangle | k_L^{\prime \prime} \rangle $ via scattering
while the $ | R \rangle | k_L^{\prime \prime} \rangle $ one is unchanged.
Hence, the polarization detection
$ \Pi ( k_R^{\prime \prime} ) $ after the scattering
effectively projects out the $ | L \rangle $ component
as recovering the initial $ | R \rangle $ state
with the success probability
$ \eta \left[ | T_{LR} (k) |^2 \right]_f = P_{\rm qm} $
where $ | T_{RL} (k) | = | T_{LR} (k) | $.
Then, if this conditional method is used to detect $ | L \rangle $
we make a substitution $ P(L) \rightarrow P(L) P_{\rm qm} $
in Eq. (\ref{eqn:P-qm}),
providing $ P_{\rm qm}^2 $ (rather than $ P_{\rm qm} $)
as the success probability of quantum memory.

In a feasible experiment to perform the present atomic quantum memory,
a sufficiently weak coherent light of $ | \alpha \rangle $
may be used as an actual single-photon source,
though the success probability becomes rather small
proportional to $ | \alpha |^2 $.
The photon detection is useful
to remove the irrelevant contribution
from the vacuum component in $ | \alpha \rangle $
as well as that from the atomic loss into the environment.
The contributions of more than one photon states in $ | \alpha \rangle $
are small enough for $ | \alpha |^2 \ll 1 $.

\section{Efficiency of quantum memory}
\label{sec:efficiency}

As seen in Eq. (\ref{eqn:phi-p2}),
for $ T_{LR} (k) \approx T_{LR} ( k_p ) $
in the vicinity of photon-pulse peak
$ | k - k_p | \lesssim \kappa_p $
with the sufficiently narrow width $ \kappa_p \ll \kappa $,
the fidelity of the present quantum memory really approaches unity
in Eq. (\ref{eqn:F-qm}):
\begin{eqnarray}
F_{\rm qm} \approx 1 \ ( \kappa_p \ll \kappa ) .
\end{eqnarray}
(In this situation with $ \kappa_p \ll \kappa $
the first Markov approximation for the usual input-output relation
will be valid essentially, where the atom-photon couplings are assumed
to be independent of the frequency $ k $
\cite{Markov-approx}.)
This high fidelity $ F_{\rm qm} $ is rather independent of
the details of the spectral profiles $ f(k) $ and $ \eta(k) $.
It is not restricted either by the finite atomic loss, cavity leakage
or detunings even in the moderate coupling regime
providing $ \left| \sin ( \phi_s ( k_p ) / 2 ) \right|^2 \sim 0.1 $,
which is in contrast with $ F_{\rm swap} $
in Eq. (\ref{eqn:F-swap-leading}).
Therefore, we find that the present method
is quite promising for implementing quantum memory
with single atoms in cavity QED;
the memory operation can be performed almost faithfully
in a reasonable range of experimental parameters.

These profitable features of the present atomic quantum memory
are really confirmed by a numerical analysis to evaluate
the fidelities in comparison with the swapping method.
We typically take the Gaussian [G] and Lorentzian [L] profiles,
which are given, respectively, by
\begin{eqnarray}
f(k) [{\rm G}]
&=& \frac{\exp [ - ( k - k_p )^2 / 2 \kappa_p^2 ]}
{\sqrt{{\sqrt \pi} \kappa_p}} e^{i ( k - k_p ) x_0} ,
\\
f(k) [{\rm L}]
&=& \frac{\sqrt{\kappa_p / \pi}}{k - k_p + i \kappa_p}
e^{i ( k - k_p ) x_0} .
\end{eqnarray}
Here, the spectral distribution is specified
with the peak position $ k_p $ and width $ \kappa_p $,
while the spatial location is given
by the factor $ e^{i ( k - k_p ) x_0} $.
The coordinate $ x_0 $ of the center is supposed
to be large enough as $ x_0 \gg 1 / \kappa_p , l $
so that the initial photon pulse is sufficiently apart from the cavity
with length $ l $.
Then, the phase shift $ \phi_s(k) $
is determined independently of $ f(k) $,
as given in Eq. (\ref{eqn:phi-s})
\cite{CLL-2004}.

\begin{figure}[t]
\begin{center}
\scalebox{0.475}{\includegraphics*[3cm,1.5cm][17cm,14cm]{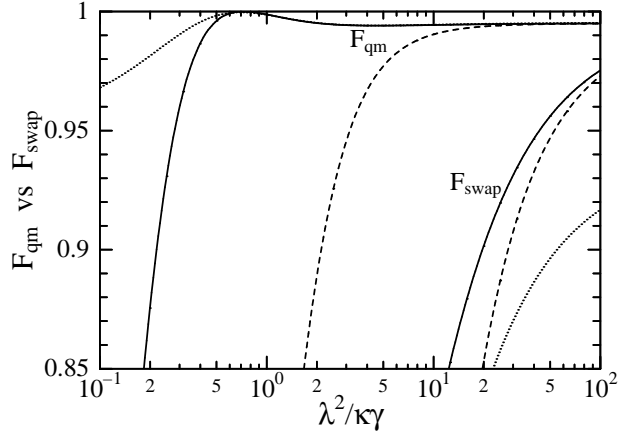}}
\caption{
$ F_{\rm qm} $ (upper) and $ F_{\rm swap} $ (lower)
are compared depending on $ \lambda^2 / \kappa \gamma $
for $ \kappa_p = 0.1 \kappa $, $ \kappa = 2 \gamma $
and (solid): $ \delta_e = 0 $, $ \delta_p = 0 $,
(dashed): $ \delta_e = 5 \gamma $, $ \delta_p = 0 $,
(dotted): $ \delta_e = 0 $, $ \delta_p = 0.5 \gamma $, respectively.
The Gaussian profile is taken here.
}
\label{FqmFswap}
\end{center}
\end{figure}

We compare in Fig. \ref{FqmFswap} the fidelities
$ F_{\rm qm} $ (upper) of the present quantum memory
and $ F_{\rm swap} $ (lower) of the swapping
depending on $ \lambda^2 / \kappa \gamma $
typically for $ \kappa_p = 0.1 \kappa $, $ \kappa = 2 \gamma $
and (solid): $ \delta_e = 0 $, $ \delta_p = 0 $,
(dashed): $ \delta_e = 5 \gamma $, $ \delta_p = 0 $,
(dotted): $ \delta_e = 0 $, $ \delta_p = 0.5 \gamma $, respectively.
The Gaussian profile is taken here.
(Precisely, $ F_{\rm qm} $ and $ F_{\rm swap}^2 $
should be compared for the storage and retrieval.)
We clearly see that this quantum memory works efficiently,
achieving the quite high fidelity $ F_{\rm qm} $
for reasonable experimental parameters
such as $ ( \gamma , \kappa , \lambda ) / 2 \pi
\approx ( 3 {\rm MHz} , 6 {\rm MHz} , 15 {\rm MHz} ) $
with $ \lambda^2 / \kappa \gamma \approx 10 $
and $ \kappa_p \approx 0.1 \kappa \approx 0.6 {\rm MHz} $
\cite{CQED,atom-CQED}.
As long as $ | \delta_e |, | \delta_p | \lesssim \gamma $
($ \sim 1 - 10 {\rm MHz} $),
the detunings do not provide significant effects on the fidelity
$ F_{\rm qm} $ for $ \kappa_p \lesssim 0.1 \kappa $
and $ \lambda^2 / \kappa \gamma \gtrsim 10 $.
As for the cavity leakage,
similar results are obtained for $ \kappa \sim ( 1 - 10 ) \gamma $.
Its optimal value is $ \kappa \sim \gamma $;
the smaller $ \kappa $ is the smaller $ \kappa_p $ should be taken
for $ \kappa_p \lesssim 0.1 \kappa $,
while the larger $ \kappa $ is the larger $ \lambda $ is required
for $ \lambda^2 / \kappa \gamma \gtrsim 10 $.
The fidelity of swapping $ F_{\rm swap} $,
on the other hand, is obviously lower than $ F_{\rm qm} $.
It is rather limited
by the finite atomic loss, cavity leakage and detunings,
as seen in Eq. (\ref{eqn:F-swap-leading}).

\begin{figure}[t]
\begin{center}
\scalebox{0.475}{\includegraphics*[3cm,1.5cm][17cm,14cm]{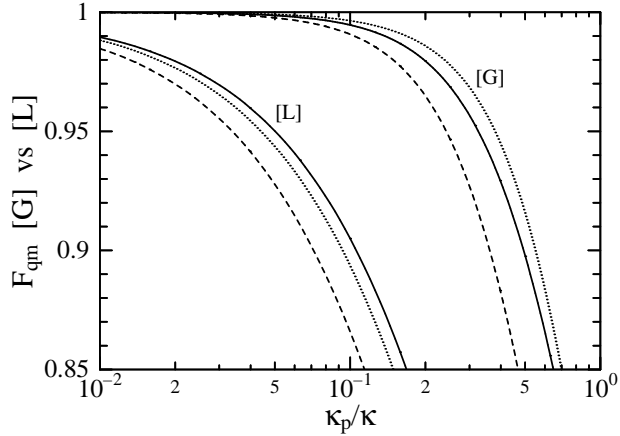}}
\caption{
$ F_{\rm qm} $ is shown depending on $ \kappa_p / \kappa $
for the Gaussian [G] and Lorentzian [L] profiles.
Here the parameters are taken
as $ \lambda^2 / \kappa \gamma = 20 $,
$ \kappa = 2 \gamma $
and (solid): $ \delta_e = 0 $, $ \delta_p = 0 $,
(dashed): $ \delta_e = 10 \gamma $, $ \delta_p = 0 $,
(dotted): $ \delta_e = 0 $, $ \delta_p = 2 \gamma $, respectively.
}
\label{FqmGL}
\end{center}
\end{figure}

We also show in Fig. \ref{FqmGL}
the fidelity of quantum memory $ F_{\rm qm} $
depending on $ \kappa_p / \kappa $
for the Gaussian [G] and Lorentzian [L] profiles.
Here the parameters are taken
as $ \lambda^2 / \kappa \gamma = 20 $,
$ \kappa = 2 \gamma $
and (solid): $ \delta_e = 0 $, $ \delta_p = 0 $,
(dashed): $ \delta_e = 10 \gamma $, $ \delta_p = 0 $,
(dotted): $ \delta_e = 0 $, $ \delta_p = 2 \gamma $, respectively.
It is observed that the Gaussian profile provides the quite high fidelity
as it has the sharp spectral peak compared to the Lorentzian profile.
We obtain for instance $ F_{\rm qm} \geq 0.999 $
with $ \kappa_p / \kappa \lesssim 0.05 $
for $ \lambda^2 / \kappa \gamma = 20 $ and $ \kappa = 2 \gamma $
if the detunings are not too large.
The detunings of atom and photon pulse do not provide significant effects,
as already seen in Fig. \ref{FqmFswap}.
It is also noticed in Fig. \ref{FqmGL}
that the asymptotic value $ F_{\rm qm} \simeq 0.995 $
for $ \lambda / \kappa \gamma \gtrsim 10 $ in Fig. \ref{FqmFswap}
is really determined by the ratio $ \kappa_p / \kappa = 0.1 $.

\begin{figure}[t]
\begin{center}
\scalebox{0.475}{\includegraphics*[3cm,1.5cm][17cm,14cm]{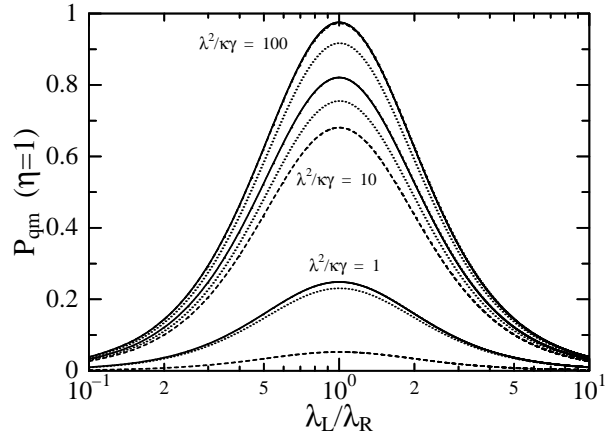}}
\caption{
$ P_{\rm qm} $ with $ \eta = 1 $ is shown
depending on $ \lambda_L / \lambda_R $
for $ \lambda^2 / \kappa \gamma $ = 1 (lower), 10 (middle), 100 (upper)
with $ \kappa_p = 0.1 \kappa $, $ \kappa = 2 \gamma $
and (solid): $ \delta_e = 0 $, $ \delta_p = 0 $,
(dashed): $ \delta_e = 5 \gamma $, $ \delta_p = 0 $,
(dotted): $ \delta_e = 0 $, $ \delta_p = 0.5 \gamma $, respectively.
The Gaussian profile is taken here.
Note that for $ \lambda^2 / \kappa \gamma = 100 $
the solid and dashed lines are almost overlapped.
}
\label{Pqm}
\end{center}
\end{figure}

The success probability of quantum memory
$ P_{\rm qm} $ with $ \eta = 1 $ is shown in Fig. \ref{Pqm}
depending on $ \lambda_L / \lambda_R $
for $ \lambda^2 / \kappa \gamma $ = 1 (lower), 10 (middle), 100 (upper)
with $ \kappa_p = 0.1 \kappa $, $ \kappa = 2 \gamma $
and (solid): $ \delta_e = 0 $, $ \delta_p = 0 $,
(dashed): $ \delta_e = 5 \gamma $, $ \delta_p = 0 $,
(dotted): $ \delta_e = 0 $, $ \delta_p = 0.5 \gamma $, respectively.
The Gaussian profile is taken here.
(Note that for $ \lambda^2 / \kappa \gamma = 100 $
the solid and dashed lines are almost overlapped.)
As seen in Eq. (\ref{eqn:P-qm}), $ P_{\rm qm} $ is proportional
to the quantum efficiency $ \eta $ of the photon detector,
and its dependence on the ratio $ \lambda_L / \lambda_R = \tan \xi $
of the dipole couplings is given by the factor $ \sin^2 2 \xi $,
indicating the trade-off for the high $ F_{\rm qm} $.
The success probability $ P_{\rm qm} $ ($ \eta = 1 $) is really optimized
to coincide with the fidelity of swapping $ F_{\rm swap} $
for $ \Lambda $-type atoms satisfying the condition
$ \lambda_L = \lambda_R $.
The effects of the detunings on $ P_{\rm qm} $
appear in the same way as $ F_{\rm swap} $.

We conclude in these calculations
that the present method for atomic quantum memory works quite efficiently,
achieving the fidelity of almost unity
with reasonable experimental parameters.

\section{An application: storage of 2-qubit entanglement}
\label{sec:2-qubit}

As an application of the quantum memory via atom-photon scattering,
we here consider storage of 2-qubit entanglement.
We prepare two atomic memories
and a polarization-entangled pair of photon pulses.
Each photon pulse is scattered with the atom inside the respective cavity.
Then, for the ideal case
of $ T_{LR} = T_{RL} = 1 $ and $ T_{LL} = T_{RR} = 0 $,
the quantum states of atom pair and photon pair,
either entangled or separable, are swapped as
\begin{eqnarray}
\left( \begin{array}{c} a_{LL} \\ a_{RR} \\
a_{LR} \\ a_{RL} \end{array} \right)_{{\rm a}^2}
\! \! \! \otimes
\left( \begin{array}{c} c_{LL} \\ c_{RR} \\
c_{LR} \\ c_{RL} \end{array} \right)_{{\rm p}^2}
\Rightarrow
\left( \begin{array}{c} c_{RR} \\ c_{LL} \\
c_{RL} \\ c_{LR} \end{array} \right)_{{\rm a}^2}
\! \! \! \otimes
\left( \begin{array}{c} a_{RR} \\ a_{LL} \\
a_{RL} \\ a_{LR} \end{array} \right)_{{\rm p}^2}
\end{eqnarray}
with the basis states
$ | L L \rangle $, $ | R R \rangle $,
$ | L R \rangle $, $ | R L \rangle $
for the atom pair ``$ {\rm a}^2 $"
and $ | k_L k_L \rangle $, $ | k_R k_R \rangle $,
$ | k_L k_R \rangle $, $ | k_R k_L \rangle $
for the photon pair ``$ {\rm p}^2 $".
This sort of 2-qubit swapping may be applied
to the storage of photonic polarization entanglement as
\begin{eqnarray}
c_{LR} | {\bar k}_L \rangle | {\bar k}_R^\prime \rangle
+ c_{RL} | {\bar k}_R \rangle | {\bar k}_L^\prime \rangle
\Rightarrow
c_{RL} | L R \rangle + c_{LR} | R L \rangle .
\end{eqnarray}

Specifically, by taking the initial state
\begin{eqnarray}
| \Phi_{\rm in} \rangle
= | R R \rangle
( c_{LR} | {\bar k}_L \rangle | {\bar k}_R^\prime \rangle
+ c_{RL} | {\bar k}_R \rangle | {\bar k}_L^\prime \rangle ) ,
\end{eqnarray}
we obtain the output state via scatterings in the cavities 1 and 2 as
\begin{eqnarray}
| \Phi_{\rm out} \rangle
&=& {\mathcal T}_1 {\mathcal T}_2 | \Phi_{\rm in} \rangle .
\end{eqnarray}
Here the $ ( k k^\prime ) $-component of $ | \Phi_{\rm out} \rangle $
is given by
\begin{eqnarray}
| \Phi_{\rm out}^{( k k^\prime )} \rangle
= | \psi_{\rm str}^{( k k^\prime )} \rangle | k_L k_L^\prime \rangle
+ | R R \rangle | \phi_{LR}^{( k k^\prime )} \rangle
\label{eqn:Phi-kkprime}
\end{eqnarray}
with
\begin{eqnarray}
| \psi_{\rm str}^{( k k^\prime )} \rangle
&=& T_{LR}(k) c_{RL} | L R \rangle
+ T_{LR}( k^\prime ) c_{LR} | R L \rangle ,
\\
| \phi_{LR}^{( k k^\prime )} \rangle
&=& T_{RR}( k^\prime )  c_{LR} | k_L k_R^\prime \rangle
+ T_{RR} (k) c_{RL} | k_R k_L^\prime \rangle .
\end{eqnarray}
Then, the fidelity $ F ( {\rm p}^2 \rightarrow {\rm a}^2 ) $
of the entanglement transfer via swapping is evaluated
by tracing over the photon states and the environment
denoted by $ | 0 \rangle \langle 0 | $.
For any choice of the initial state it is bounded as
\begin{eqnarray}
| [ T_{LR} ]_f |^2 \leq
F ( {\rm p}^2 \rightarrow {\rm a}^2 )
\leq [ | T_{LR} |^2 ]_f = \sin^2 2 \xi F_{\rm swap} ,
\end{eqnarray}
where $ | [ T_{LR} ]_f |^2 \leq [ | T_{LR} |^2 ]_f $
for $ F_{\rm qm} \leq 1 $ in Eq. (\ref{eqn:F-qm}) is considered.
For the specific case of $ \lambda_L = \lambda_R $ ($ \sin 2 \xi = 1 $),
the fidelity is optimally given by $ F_{\rm swap} $ of the 1-qubit swapping
in Eq. (\ref{eqn:F-swap}), while it is rather below unity
for $ \lambda_L \not= \lambda_R $.

Alternatively, as done in Sec. \ref{sec:memory-operation},
we can make actively the photon detection
$ \Pi ( k_L ) \otimes \Pi ( k_L^\prime ) $
on the output state in Eq. (\ref{eqn:Phi-kkprime})
to remove the undesired term
$ | R R \rangle | \phi_{LR}^{( k k^\prime )} \rangle $,
so that the trade-off of the success probability
is made to obtain the high fidelity.
Then, the transfer of the photonic entanglement
to the atomic memories can be implemented almost faithfully.
The fidelity of this 2-qubit entanglement memory
is calculated to be identical to that of the 1-qubit memory
in Eq. (\ref{eqn:F-qm}):
\begin{eqnarray}
F_{\rm 2qem} = F_{\rm qm} \approx 1 \ ( \kappa_p \ll \kappa ) .
\end{eqnarray}
Reversely, the entanglement stored in the pair of atomic quantum memories
is retrieved by injecting single-photon pulses
(which may be separable each other) to the atomic memories.
In a feasible experiment for this entanglement transfer,
a photon pair from the type-II parametric down-conversion
may be used as the input polarization-entangled qubit.

\section{Summary}
\label{sec:summary}
In summary, we have investigated a scheme of atomic quantum memory
to store photonic qubits of polarization in cavity QED.
The swapping between photonic and atomic qubits
can be made via scattering in optical cavities.
This swapping operates limitedly in the strong coupling regime
for $ \Lambda $-type atoms with equal dipole couplings.
By extending this scheme of atom-photon scattering in cavity QED,
we have presented a more feasible and efficient method
to implement the quantum memory operation
combined with projective measurement.
This method works without requiring such a condition
on the dipole couplings of $ \Lambda $-type atoms.
The fidelity is significantly higher than that of the swapping,
and even in the moderate coupling regime
it reaches almost unity by narrowing sufficiently the photon-pulse spectrum.
This high performance is rather unaffected
by the atomic loss, cavity leakage or detunings,
while a trade-off is paid in the success probability
for projective measurement.

\bigskip

\begin{flushleft}
{\bf Acknowledgments}
\end{flushleft}

\bigskip

The authors would like to thank M. Kitano, A. Kitagawa, and K. Ogure
for valuable suggestions and comments.
This work is supported by International Communications Foundation (ICF).



\end{document}